# Against the disappearance of spacetime in quantum gravity

Michael Esfeld

University of Lausanne, Department of Philosophy

CH-1015 Lausanne, Switzerland

Michael-Andreas.Esfeld@unil.ch, www.michaelesfeld.com

## Abstract

This paper argues against the proposal to draw from current research into a physical theory of quantum gravity the ontological conclusion that spacetime or spatiotemporal relations are not fundamental. As things stand, the status of this proposal is like the one of all the other claims about radical changes in ontology that were made during the development of quantum mechanics and quantum field theory. However, none of these claims held up to scrutiny as a consequence of the physics once the theory was established and a serious discussion about its ontology had begun. Furthermore, the paper argues that if spacetime is to be recovered through a functionalist procedure in a theory that admits no fundamental spacetime, standard functionalism cannot serve as a model: all the known functional definitions are definitions in terms of a causal role for the motion of physical objects and hence presuppose spatiotemporal relations.

## 1. The functionalization of spacetime

It is widely alleged that when it comes to the physics of quantum gravity, spacetime turns out to be no longer fundamental (see e.g. Rovelli 2004, ch. 10, Kiefer 2007, ch. 5, Oriti 2014, Crowther 2016, Wüthrich 2017). The claim is not only that there is no fundamental spacetime substance, but also that spatiotemporal relations are not fundamental. Thus, the claim is not that, as there is a shift from Euclidean to Riemannian geometry in the transition from Newtonian gravitation to Einstein's general theory of relativity, so there may be another such shift in geometry in the transition to a quantum theory of gravity. The claim is much more radical, namely that after that transition, there will be nothing left in fundamental physics that is like the spacetime or the spatiotemporal relations with which we are familiar. Thus, this claim does not concern string theory and its unification project: on string theory, there is a fundamental spacetime that even serves as background space. It has more dimensions than four, but, still, string theory is distinct from positions that reject the fundamentality of spacetime. Prominent examples of approaches to a physical theory of quantum gravity on the basis of which this radical conclusion is drawn are loop quantum gravity and causal set theory.

This paper is an argument against this conclusion. *In nuce*, the argument to be developed in the following two sections is that this claim is on a par with all the other claims about radical changes in ontology that have been made during the development of quantum mechanics and quantum field theory. None of these claims held up to scrutiny as a consequence of the physics once the theory was established and a serious discussion about its ontology had begun. Furthermore, the argument of this paper is that as things stand, there is no other worked out functionalism than the one in terms of causal or functional roles for the motion of spatiotemporally related objects. This functionalism thus presupposes spacetime (at least in the guise of fundamental spatiotemporal relations). It can hence not serve as a model for a research programme whose aim is to show how spacetime and spatiotemporally related objects enter into a theory that does not admit a fundamental spacetime.

The claim that there are no fundamental spatiotemporal relations could be true. The argument of this paper is not about an alleged empirical incoherence of physical theories that do not admit fundamental spatiotemporal relations. It is difficult to sustain any *a priori* claims about empirical incoherence in a science based or naturalized metaphysics. The most promising strategy to show how a theory that does not admit fundamental spatiotemporal relations can be true is functionalism about spacetime (see in particular Lam and Wüthrich 2018). The functionalism at issue here goes beyond the functionalism about spacetime that Knox advocates in her recently published paper (2018). The functionalism defended in this paper is a functionalism about geometry only. According to Knox (2018, p. 5), "… spacetime is whatever serves to define a structure of inertial frames, where inertial frames are those in whose coordinates the laws governing interactions take a simple form (…), and with respect to which free bodies move with constant velocity". Inertial frames obviously presuppose objects that move and that hence are spatiotemporally related. This functionalism concerns only the geometry, which, according to this view, enters into a physical theory through its causal or functional role for the motion of objects. It is thus akin to standard functionalism.

Consider functionalism in the special sciences. For instance, there is no fundamental water stuff in the world. Nevertheless, there is water in the world, namely things that fulfil the functional role of appearing odourless, colourless, being thirst-quenching through the change in the motion of the parts of our bodies that they bring about, and these are configurations of $H_2O$ molecules (for the sake of the example, let us leave aside here the arguments according to which "water" is a rigid designator). By the same token, there is no *élan vital*; but there are living organisms in the world. The functional role that defines what it is to be alive in terms of characteristic motions such as reproduction and adaptation to the environment is realized by certain configurations of molecules. Furthermore, there arguably are no *sui generis* minds; but there are mental states defined by certain functional roles for the behaviour and thus the bodily motions of persons. These roles are fulfilled by certain configurations of neurons. In all these cases, something goes – the primitive water stuff (think of the four elements earth, water, air, fire), the primitive life, the primitive mind (or consciousness). But water, life, the mind remains, since it is functionalized. What the functional definition in terms of a causal role for the motion of bodies captures arguably is all there is to water, life, the mind.

There is a very good motivation for pursuing such a functionalist project. Consider how Jackson (1994, p. 25) describes the task of metaphysics:

Metaphysicians seek a comprehensive account of some subject matter – the mind, the semantic, or, most ambitiously, everything – in terms of a limited number of more or less basic notions. … But if metaphysics seeks comprehension in terms of limited ingredients, it is continually going to be faced with the problem of location. Because the ingredients *are* limited, some putative features of the world are not going to appear explicitly in the story. The question then will be
whether they, nevertheless, figure implicitly in the story.

Functionalism is the answer to that question: water, life, minds, etc. do not appear explicitly in the story of basic physical objects and laws for their motion. But they figure implicitly in that story: some configurations of basic physical objects moving in a particular way *are* water, others *are* living beings, and yet others *are* minds because they fulfil the functional roles that characterize water, life, or minds in terms of the behaviour and thus the motion of bodies. These configurations may be configurations of point particles, field configurations or configurations of events, whatever may come out as the basic, spatiotemporally related physical objects according to classical mechanics, the standard model of elementary particle physics or general relativity theory.

Functionalism is a reductionist programme in the following sense: the tokens of water, living beings, minds, etc. that exist in the universe are identified with certain configurations of in the last resort basic physical objects and their characteristic motions. If a complete physical description of the configurations of basic physical objects in the universe and their motions were given together with the functional definitions, that description would entail which of these configurations are water, which ones are living beings, which ones are minds, etc. (although one may have reservations about Jackson's (1994) claim that this is a matter of *a priori* entailment). Hence, there is no question here of emergence in the sense of something new being added to the ontology. It is true that only some configurations of basic physical objects are water, living beings, or minds. These configurations exist only at particular times and only in certain regions of the universe. Nonetheless, since the configurations in question *are* water, living beings, or minds, no additional ontological commitments that go beyond the commitments to basic physical objects, their configurations and motions are called for. In brief, functionalism allows us to locate features of the world in an ontology that is conceived only in terms of a few basic and general notions of a fundamental physical theory of the universe.

This motivation for functionalism may apply to spacetime as well. If spacetime, including spatiotemporal relations, is not fundamental, it better enters the ontology through functionalization, since there is no question of eliminating it. In that way, it can be shown how it is located in something that is not spatiotemporal. The term "location" is used by Jackson (1994) in the quotation above in a metaphorical sense. It is not necessarily tied to spatiotemporal location – although in the standard examples of functionalism, there is location in the literal sense.

There is no *a priori* reason that obliges us to admit spacetime – or spatiotemporal relations – to the ontology being defined in terms of the fundamental and general notions. Spacetime may be open to functionalization as well. Something about spacetime then has to go, namely the idea that it is a substance, or that at least the network of spatiotemporal relations is a network of irreducible relations. Nonetheless, it may be possible to functionalize the essential features of spacetime that are relevant for the account of what we know about spacetime and spatiotemporally related objects. In this case, then, say, some particular spin networks (if loop gravity turns out to be the correct theory), or some particular causal sets (if causal set theory turns out to be the correct theory), *are* relativistic spacetimes in the sense that they fulfil the functional role that relativistic spacetime plays in the accomplished scientific image of the world that includes quantum gravity.

However, one has to be clear about the task that this project faces: by contrast to the well understood standard examples of the functionalization of something, what has to be provided in this case is a functional definition of something in terms of a functional role that is not a role which is defined in terms of the motion of spatiotemporal objects, but a functional role for the evolution of, for instance, spin networks or causal sets. To put it in other words, in the terms of Lewis (1970, 1972), some other original or old (that is, basic) vocabulary has in this case to be provided that does not rely on the familiar old vocabulary of spatiotemporal relations between physical objects (or spatial relations and their change) and with respect to which functional roles that objects standing in these relations fulfil are defined. In particular, in this case, we need another world-making relation than the spatial or spatiotemporal one. As things stand, spatial or spatiotemporal relations are world-making because all and only those objects that are spatially or spatiotemporally related belong to the same world; objects that are not thus related belong to different possible worlds (cf. Lewis 1986, ch. 1.6).

Again, functionalizing spacetime is a reductionist project in the mentioned sense. Although the issue is the emergence of spacetime from something that is fundamentally not spatiotemporal, this is not an emergence as opposed to reduction: some spin networks, or some causal sets, then *are* relativistic spacetimes. If the complete description of the spin networks or causal sets of the universe were given, that description would *entail* which spin networks, or causal sets, are relativistic spacetimes. It is only that the concepts characterizing relativistic spacetimes do not figure in the description of the spin networks or causal sets and that only some spin networks or causal sets are spacetimes. But this is just what applies to functionalism in general as a solution to the problem of location. Location is a reductionist affair: identity as far as ontology is concerned, and entailment as regards epistemology. There is nothing *a priori* incoherent with such a functionalist project, merely because this time it tackles spacetime, instead of water, life, or the mind. The issue is whether there is, as things stand in physics, any cogent reason to believe that such a project has or is likely to have the same status that it enjoys in chemistry, biology or psychology. Taking into account the standard for ontology in the established areas of quantum physics (quantum mechanics and quantum field theory) suggests that there is no such reason.

## 2.    The case of quantum mechanics

Consider non-relativistic quantum mechanics. The algebra of operators in a Hilbert space is not Boolean. Does it follow from this that classical logic no longer is appropriate for the description of the world? The Heisenberg uncertainty relations exclude that it is possible to measure both the position and the momentum of a quantum system with arbitrary precision at the same time. Does it follow from this that quantum systems do not have definite positions and that they do not move on continuous trajectories? It is in general only possible to make statistical predictions of measurement outcome distributions, but not deterministic predictions of single measurement outcomes. Does it follow from this that the laws to which quantum systems are subject are not deterministic and that probabilities in quantum mechanics have a more fundamental status than in classical statistical mechanics (where they are derived from the deterministic laws of classical mechanics plus an appropriate probability measure)?

No one who is familiar with the debate about the ontology of quantum mechanics would answer these questions in the affirmative. The reason is not that, at the end of the day, it may not be established that quantum mechanics requires amending classical logic, that quantum systems usually do not have definite positions and that they do not move on continuous trajectories and that probabilities are irreducible in quantum mechanics. The point at issue is that these claims do not follow from the physics of quantum mechanics *per se*. If one is to make a case for these claims, extensive philosophical argument is required. The reason for this is threefold:

1) Generally speaking, there is no possibility to read ontological conclusions off from the mathematical structures that physical theories employ. Any reasoning that leads to ontological conclusions requires ontological premises.

2) The above mentioned claims, notably the second one, run into the measurement problem (see Maudlin 1995 for the by now standard formulation of this problem): if quantum systems do in general not have definite positions, how does it come about that they are found in definite positions when measured? How do they get from the preparation at a determinate position to the measurement at a determinate position? Note that all measurement records in physics are records of positions of discrete objects, as Bell (2004, p. 166) among others pointed out.

3) There are formulations of quantum mechanics in which none of the three mentioned claims is true. These formulations are coherent, they propose a solution to the measurement problem, and they yield the correct statistical predictions. Thus, any attempt to establish one of the above mentioned claims has to give reasons why it is a better candidate for the truth about the world than these formulations.

In sum, the standard for ontology in quantum mechanics is the solution to the measurement problem (cf. Albert 1992). Any suggestions for ontological consequences from quantum mechanics have to spell out how they come to terms with this problem.

Consider Bohmian mechanics as one formulation of quantum mechanics in which none of the three mentioned claims comes out true as a consequence of the way in which this theory solves the measurement problem. This theory will prove to be particularly instructive when we return to the issue of spacetime in quantum gravity. On the way in which Dürr, Goldstein and Zanghì (2013) set up Bohmian mechanics, the ontology (or primitive ontology, as they call it) is point particles that always have a position and that move on continuous trajectories in physical space. This gives us an account of the individual physical processes that happen in nature that provides for measurement outcomes and that avoids the well known paradoxes such as Schrödinger's cat.

The particle trajectories – that is, the velocities of the particles at any time given their positions at that time – are fixed by a law of motion, the so called guiding equation, by means of the wave function. The wave function, in turn, develops according to the Schrödinger equation. These laws are joined with a probability measure, defining what is known as quantum equilibrium, from which then follows Born's rule for the predictions of measurement outcome statistics on ensembles of systems in the universe that are prepared in the same manner. The operators or observables of quantum mechanics (including spin) then are construed as a means to make statistical predictions about how the particles move in certain experimental contexts (see Lazarovici et al. 2018). Hence, they are not properties of anything. The only properties of the particles are their position and change of position (motion).

If position and its change is the only primitive property of the particles, then Bohmian mechanics is committed to functionalism already for variables such as mass and charge. They have to be located in an ontology that admits only naked particles and their motion as primitive. They can thus be located by introducing them in terms of a functional role for the motion of the particles. Such a view can arguably already be applied to mass in Newtonian mechanics, as for instance Mach (1919, p. 241) pointed out in his remark "The true definition of mass can be deduced only from the dynamical relations of bodies". To put it in a nutshell, some particles are electrons because they move electronwise. What it means to move electronwise is captured in terms of the functional role for particle motion that defines the parameter of charge. The particles are not in an intrinsic or primitive way electrons (see Dickson 2000 and Esfeld et al. 2017 as well as the identity based version of Bohmian mechanics set out in Goldstein et al. 2005a, 2005b). Against this background, Bohmian mechanics can then buy into the standard functionalism for the special sciences with the functional definitions being in the last resort ones in terms of functional roles for particle motion.

The probabilities in Bohmian quantum mechanics have the same status as the probabilities in classical statistical mechanics: they follow from a deterministic law via an appropriate probability measure. In general, the ontology in the sense of primitive ontology of Bohmian mechanics is the same as the one of classical mechanics: point particles moving on continuous trajectories that are determined by a deterministic law. The only difference is on the level of the law: the trajectories are determined by the wave function. The wave function is defined on configuration space (for $N$ particles, the configuration space has $3N$ dimensions so that each point of configuration space represents a possible configuration of $N$ particles in three-dimensional physical space). This has the consequence that the evolution of the wave function in configuration space represents the motion of the particles in physical space as being correlated independently of the spatial distance among these particles. This non-locality is necessary in Bohm's quantum theory in order to accommodate Bell's theorem and the subsequent experiments (see Bell 2004, in particular chs. 2 and 7). Any quantum theory that is about objects in physical space has to admit a non-locality in the sense of correlated behaviour of these objects independently of their spatial or spatiotemporal distance in physical space.

Consequently, on Bohmian mechanics, there never are superpositions of anything in physical space. There only is correlated particle motion, with these correlations being determined by the entanglement of the wave function. Hence, there is no paradox of how quantum objects can show up at precise locations when measured, although they do in general not have positions in space. There are even arguments developed in the literature to the effect that Bohmian mechanics provides the best solution to the measurement problem in comparison to rival accounts (see e.g. Bricmont 2016, chs. 5-6, and Esfeld 2014).

However, Bohmian mechanics is in general not useful as a calculatory device for working physicists. There often is no point in calculating Bohmian particle trajectories. The reason is the high sensitivity of quantum systems to slight variations in the initial conditions together with the fact that the initial conditions cannot be fixed with arbitrary precision. The rationale of Bohmian mechanics is to solve the measurement problem – in other words, to provide a full and coherent answer to the question of what nature is like if quantum mechanics is true. Such a theory usually comes later than the formalism for working physicists (after a first attempt by de Broglie 1928, the theory had been developed for many particles by Bohm 1952).

Generally speaking, the worked out solutions to the measurement problem fall into two camps: on the one hand, there is the primitive ontology answer in terms of localized objects in three-dimensional space or four-dimensional spacetime, known as "local beables" (that neologism goes back to Bell 2004, ch. 7). The wave function provides for the dynamics of these objects, but it is not a physical object in addition to the "local beables". Bohmian mechanics is the most prominent example of this answer.

On the other hand, there is the wave function monism answer that regards the quantum wave function as the only physical object and that seeks to account for the experimental evidence on this basis. This idea goes back to Everett (1957). One version of this camp is the configuration space realism developed notably by Albert (1996, 2015, chs. 6-7). This version is relevant for present purposes, because it does not regard ordinary spacetime or ordinary spatiotemporal relations as fundamental. It sets out to give a functional account of ordinary spatiotemporal objects on the basis of what it admits as primitive. It thereby provides the up to now only example of a worked out ontology for an existing physical theory that does not endorse spacetime as fundamental and that employs functionalism to achieve empirical adequacy (that link is also discussed in Lam and Wüthrich 2018, pp. 41-42).

According to this stance, the ontology is the universal wave function, which is a field on a very high dimensional space that is usually known as the configuration space of the universe. Strictly speaking, however, on this view, this is not a configuration space, since there are no configurations of anything that its points represent. This space is fundamental. Configuration space is the physical space. There is a matter field on this space in the guise of the universal wave function. On this view, none of the three claims mentioned at the beginning of this section and suggested by a superficial consideration of quantum mechanics is true either (although both this view as well as the primitive ontology view can accommodate a dynamics of wave function collapse, which is probabilistic instead of deterministic; however, this dynamics then is not based on the statistical predictions of measurement outcomes, but is motivated by a solution to the measurement problem in terms of wave function collapse and in fact yields statistical predictions that deviate in some cases from the ones obtained by textbook quantum mechanics).

The procedure to account for ordinary objects and the experimental evidence on this basis then is exactly the one described as the motivation for functionalism in the first section above: one gives a functional definition of ordinary objects including the experimental evidence and locates something that fulfils that functional definition in the evolution of the universal wave function on configuration space. That is to say: some degrees of freedom in the wave function field in some regions of configuration space *are* tigers, cats, measurement records in laboratories, etc., functionally defined. The claim thus is that a physical realization of these functionally defined things in our universe does not require composition by spatiotemporally related objects ("spatiotemporally related" understood in the ordinary sense of a low-dimensional space).

So it seems that we have two rival accounts of the experimental results of quantum mechanics that both rely on functionalism. Both endorse functionalism with respect to the special sciences and common sense. Both postulate theoretical entities as the realizers of those functional roles in the universe – configurations of point particles, a wave function field on configuration space. None of these is or will ever be observable. However, there is an important difference between these two functionalist projects. For any functionalist project to succeed, the definition of the features that come up for functionalization cannot just be a definition in terms of some functional role or other, but it has to be a role *for – and thus exercized by – the behaviour of those entities that make up the fundamental ontology*. Otherwise, the project to show how these features are located in the fundamental ontology and therefore figure implicitly in the story told in the terms of the fundamental ontology would be a non-starter. It would then not be possible to figure out the realizers of these functional roles.

To appreciate this crucial point, consider again functionalism in the philosophy of mind. This position is the successor of behaviourism (cf. Lewis 1966). The definition of mental states in terms of a causal or functional role includes causal relations to other mental states (in contrast to behaviourism); but in the end, it has to be a definition of mental states in terms of their effects on bodily motions, which finally are motions of the molecules that compose the body and which have a molecular cause in the brain. Otherwise, the project to locate mental states in the scientific image of the world would be doomed from the start (cf. Lewis 1972).

Consider for illustration the contrast between what is known as a right wing Sellarsian functionalism of mental states in terms of their biological function, which in the end is a functional role for molecular motion (e.g. Millikan 1993), versus what is known as a left wing Sellarsian functionalism of mental states in terms of *sui generis* normative roles, that is, roles that stay within the normative space of reasons (e.g. Brandom 1994) (both these versions of functionalism go back to Sellars 1956). In the former case, the functional roles are defined in terms of effects on the motion of three-dimensional physical objects. Consequently, the mental states are in the last resort located in the particle configuration of the universe – more precisely, particle configurations in the brain. The latter also is a case of functionalism. However, there is no such location, since the functional roles are not conceived as roles for bodily motion, but as irreducibly normative roles. By way of consequence, the ontology of the mental as defined by normative roles remains in limbo, because it is unclear how these functional roles are realized. Of course, there is no question of any sort of normative functionalism when it comes to quantum gravity. But this example highlights that one cannot simply endorse functionalism with respect to the features that do not figure explicitly in the fundamental ontology and take for granted that the functional roles are realized by and thus located in what is admitted to the ontological basis. The functional definitions have to be adapted to the fundamental ontology: they have to be definitions in terms of causal roles for the behaviour of the entities in the ontological basis in order to be in the position to show how the functionally defined features enter into the ontology.

To further stress this point, consider one of the earliest statements of atomism by Democritus:

> ... substances infinite in number and indestructible, and moreover without action or affection, travel scattered about in the void. When they encounter each other, collide, or become entangled, collections of them appear as water or fire, plant or man. (fragment Diels-Kranz 68 A57, quoted from Graham 2010, p. 537).

What makes atomism attractive comes out clearly in this quotation: on the one hand, it is a proposal for a theory about what there is in the universe that is both most parsimonious and most general. On the other hand, it offers a clear and simple explanation of the realm of the objects that are accessible to us in perception. Any such object is composed of a finite number of discrete, pointlike particles. All the differences between these objects – at a time as well as in time – are accounted for in terms of the spatial configuration of these particles and its change. Consequently, there is a direct link proposed here from the fundamental ontology to the perceptible macroscopic objects in terms of composition and change of composition, for which spatial relations are crucial. That link is independent of any particular physical theory (particles may be replaced with field configurations, or events; the dynamics may be local, or non-local; it may be deterministic, or stochastic, etc.). It provides a framework within which all the notions that figure in the description of the natural world and that are not defined on the levels of atoms can then be introduced in terms of their causal or functional role for the change in the atomic composition.

The conclusion hence is this one: *one cannot take the functionalism with respect to the special sciences, including common sense and ordinary macroscopic objects, as model and abandon the functional definitions of these things in terms of causal roles for the motion of spatially or spatiotemporally related physical objects*. This functionalism provides no clue as to how these things could be located in something else than in configurations of such objects. This is the reason for the reservations that one may have as to whether wave function monism has accomplished a functionalist account of ordinary macroscopic objects including the experimental evidence and thereby solved the measurement problem. To establish such a functionalism, *new* functional definitions of these things would be required in terms of their effects on the evolution of the universal wave function in configuration space. To put it differently, the functionalist programme with which we are familiar works in terms of functional roles for bodily motions, which are finally molecular or particle motions (or the ones of field configurations or events in ordinary spacetime). It thus takes the composition of these bodies by molecules and finally point particles (or field configurations, or events) for granted and thereby is committed to ordinary spacetime, viz. spatiotemporal relations.

Coming back to functionalism about spacetime in quantum gravity, taking the debate about the ontology of and functionalism in quantum mechanics into consideration leads to two conclusions:

1) The standard for ontology are not suggestions on the basis of an algorithm to calculate measurement outcome statistics. The standard is the solution to the measurement problem, that is, an answer to the question how the features that are peculiar to the quantum formalism (such as the superposition principle) relate to empirical reality.

2) Existing versions of functionalism in quantum mechanics, including a functionalism that does not regard spacetime as fundamental, but the universal wave function in configuration space, provide no template as to how a functionalism with respect to spacetime in quantum gravity could be worked out. The reason is that we do not have functional definitions at our disposal that are not definitions with respect to the behaviour of spatiotemporal objects.

These two aspects are related: it is not clear how a functionalism that echews functional or causal roles for the behaviour of spatiotemporal objects could accommodate measurement outcomes.

*3.　　The case of relativistic physics: quantum field theory and general relativity*

When it comes to quantum field theory, the situation is similar to the one in quantum mechanics. The formalism to calculate measurement outcome statistics works with a varying number of particles, and there is experimental evidence of particles popping up and disappearing. Does it follow from this that particles are not fundamental? No, it does not, for the same three reasons as it does not follow, for instance, from the Heisenberg uncertainty relations that there are no particles moving on continuous trajectories. The measurement problem hits quantum field theory in the same way as quantum mechanics (see notably Barrett 2014). The types of solutions are the same as in quantum mechanics. In particular, there is a Bohmian solution available that is based on an ontology of particles. The first such worked out proposal was the one of an ontology with a varying number of particles with a probabilistic law of motion (see Bell 2004, ch. 19, and Dürr et al. 2005). More recently, also an ontology of a very large, but finite number of permanent particles (the so called Dirac sea) that move according to a deterministic law of motion (guiding equation) has been set out for quantum field theory. Also from this ontology follows the Fock space formalism to calculate measurement outcome statistics in terms of particle creation and annihilation operators via an appropriate probability measure (see Colin and Struyve 2007, Deckert et al. 2018). Thus, again, there is no question of endorsing far reaching ontological consequences, such as abandoning the idea of there being (permanent) particles in nature, on the basis of an operator formalism to calculate measurement outcome statistics. The question is which proposal for an ontology of quantum field theory offers the overall best explanation of the evidence and, in particular, the most convincing solution to the measurement problem. A serious debate about the answers to that question has only just begun.

Turning to general relativity theory, there is no such debate about the ontology of this physics as there is about the ontology of quantum physics. Nonetheless, recent research has made clear that this physics has a dual nature: one can either set up this physics in the guise of a dynamics with a metric of a four-dimensional spacetime in which there is no privileged foliation of this spacetime and thus no objective simultaneity, but there then is an absolute scale that defines the spatiotemporal intervals between events. Or one can set up this physics in terms of configurations that are defined by their shape only. In this case, one requires well-defined three-dimensional spatial configurations, but uses only scale-invariant quantities (see Gomes et al. 2011, Gomes and Koslowski 2013 as well as Gryb and Thébault 2016, in particular pp. 692-697). Developing the shape dynamics that goes back to the work of Barbour and collaborators since the 1970s (e.g. Barbour and Bertotti 1982, Barbour 2012) as a theory of gravitation is an in principle possibility, not a practical advice for carrying out concrete calculations. Nonetheless, the issue of two different versions of a theory of gravitation cannot be settled by observation: both yield the same particle trajectories given appropriate restrictions. One can neither observe absolute scales nor absolute simultaneity. Hence, again, far reaching ontological consequences cannot be built on the fact that working physicists use the formalism of a four-dimensional, curved spacetime with no privileged foliation. The issue again is a matter of philosophical debate about which ontology constitutes the overall best explanation.

*4.    Conclusion*

Coming back to the alleged disappearance of spacetime in quantum gravity, we can make three points on the basis of the preceding considerations with respect to the ontology of quantum mechanics and quantum field theory as well as general relativity theory. (1) In the first place, the spacetime functionalism in quantum gravity faces the same problem as the functionalism in wave function monism: as things stand, there are no functional definitions provided that locate spacetime or spatiotemporal relations in causal roles for the behaviour of non-spatiotemporal spin networks or causal sets. The most advanced paper in that respect, Lam and Wüthrich (2018) does not spell out such definitions. Again, the usual functional definitions provided in standard functionalism with respect to the special sciences, or with respect to dynamical parameters such as mass and charge, are not applicable in this case. They are construed in terms of causal roles for the behaviour in the sense of the spatiotemporal evolution of something.

(2) Most importantly, the approaches to quantum gravity that allegedly entail that spacetime or spatiotemporal relations are not fundamental are approaches that, as things stand, do not yield any empirical predictions. Overcoming this lack is not just a matter of further pursuing these approaches as they are conceived up to now. The principled reason for the lack of empirical predictions is that, as things stand, these approaches do not take matter into consideration. However, all empirical predictions are about the behaviour – that is, the motion – of matter. Recall that, for instance, even in the case of the gravitational waves detected by LIGO (Laser Interferometer Gravitational-Wave Observatory) in 2016 that confirm general relativity theory, what is observed and what is the evidence for gravitational waves consists in the motion of material objects.

Hence, the measurement problem that any quantum formalism faces due to the superpositions in the wave function does not even come into the view of these approaches, not to mention a solution to that problem. However, the measurement problem is the standard for ontology in any quantum theory, because it provides the link between the formalism and the physical reality. The obligation to solve this problem motivates pursuing a primitive of ontology of material objects for which at least spatiotemporal relations are fundamental, that provides an ontology for quantum theories in the areas of mechanics and the standard model of elementary particles as well as gravity and that gives an account of the individual processes in nature that lead to measurement outcomes.

To illustrate this assessment in terms of one obvious example, consider the fact that in the Wheeler-deWitt equation figures a stationary wave function. That is to say, by contrast to the Schrödinger equation in quantum mechanics and the Dirac equation in quantum field theory, in the corresponding equation in loop quantum gravity, the wave function of the universe no longer develops in time. However, from this fact follows a problem of time – or the entitlement to the conclusion of the disappearance of time on the fundamental level – if and only if one presupposes wave function monism. In other words, this far reaching ontological conclusion follows only if one adds a controversial ontological premise. If one pursues a primitive ontology approach in order to solve the measurement problem, the wave function and the equation for it only have a dynamical role as being the central element in providing a dynamics for the evolution of the configuration of the primitive ontology. A stationary wave function can do so in the same manner as one that develops in time. Furthermore, if it turned out that the wave function of the universe were stationary, this would strengthen the claim made in the context of primitive ontology theories that it is nomological by contrast to being a physical object (see Dürr, Goldstein and Zanghì 2013, ch. 12, for that claim).

(3) Lam and Wüthrich (2018, p. 41) write: "Most importantly, the QM context allows for spacetime-based ontologies that constitute alternatives to wave function realism in a way that is not straightforwardly available or legitimate in the context of QG". This statement is simply not true. Spacetime-based ontologies, which add a primitive ontology of spatiotemporally related material objects to the wave function, are legitimate in the context of quantum gravity in the same way as in the context of quantum field theory or the context of quantum mechanics, namely as reply to the question of how the formalism relates to physical reality and, in particular, as solution to the measurement problem. They are also available in the context of quantum gravity. For instance, a Bohmian primitive ontology for quantum gravity is pursued in Goldstein and Teufel (2001), Vassallo and Esfeld (2014) and Pinto-Neto and Struyve (2018). A primitive ontology of GRW flashes is pursued in Tilloy (2018) (see also Okon and Sudarsky 2014 for wave function collapse models in quantum gravity). Furthermore, even if one considers only spacetime, leaving aside matter, shape dynamics is available also in quantum gravity as an approach that bases itself on fundamental spatial relations (see notably Barbour, Koslowski and Mercati 2013 and Gryb and Thébault 2016).

In conclusion, there is no reason to suppose that the status of the claim that there is no fundamental spacetime is in any way superior to the status of the claims about changing logic to make sense of quantum mechanics, that quantum objects do not have positions and trajectories, that there are no (permanent) particles in the ontology of quantum field theory, etc. One may make a sound case for such claims. But the case has to be made by arguing that, in contrast to rival approaches that fully match the algorithm to calculate measurement outcome statistics and that do not make any such claims, endorsing such far reaching ontological consequences leads to the overall best explanation and, notably, the best solution to the measurement problem. There is no question of it being possible to establish such ontological claims directly on the basis of certain formalisms, as if there were no alternative to them for anyone who takes the physics seriously. As regards quantum gravity, we are still quite far from being able to engage in a serious discussion about which approach provides the overall best explanation and, in particular, the best solution to the measurement problem. Consequently, as things stand, it is reasonable to recommend caution about proposing far reaching ontological consequences such as the disappearance of spacetime or fundamental spatiotemporal relations.

*Acknowledgements*: I'm grateful to the organizers and the participants of the workshop on "Spacetime functionalism" in Geneva in March 2018 for their feedback as well as to two anonymous referees for very helpful comments on the first version of this paper.

## References

Albert, David Z. (1992): *Quantum mechanics and experience*. Cambridge (Massachusetts): Harvard University Press.

Albert, David Z. (1996): "Elementary quantum metaphysics". In: J. T. Cushing, A. Fine and S. Goldstein (eds.): *Bohmian mechanics and quantum theory: An appraisal*. Dordrecht: Kluwer. Pp. 277-284.

Albert, David Z. (2015): *After physics*. Cambridge (Massachusetts): Harvard University Press.

Barbour, Julian B. (2012): "Shape dynamics. An introduction". In: F. Finster, O. Mueller, M. Nardmann, J. Tolksdorf and E. Zeidler (eds.): *Quantum field theory and gravity*. Basel: Birkhäuser. Pp. 257-297.

Barbour, Julian B. and Bertotti, Bruno (1982): "Mach's principle and the structure of dynamical theories". *Proceedings of the Royal Society A* 382, pp. 295-306.

Barbour, Julian B., Koslowski, Tim and Mercati, Flavio (2013): "The solution to the problem of time in shape dynamics". *Classical and Quantum Gravity* 31, no. 15.

Barrett, Jeffrey A. (2014): "Entanglement and disentanglement in relativistic quantum mechanics". *Studies in History and Philosophy of Modern Physics* 47, pp. 168-174.

Bell, John S. (2004): *Speakable and unspeakable in quantum mechanics*. Cambridge: Cambridge University Press. Second edition.

Bohm, David (1952): "A suggested interpretation of the quantum theory in terms of 'hidden' variables. I and II". *Physical Review* 85, pp. 166-179, 180-193.

Brandom, Robert B. (1994): *Making it explicit. Reasoning, representing, and discursive commitment*. Cambridge (Massachusetts): Harvard University Press.

Bricmont, Jean (2016): *Making sense of quantum mechanics*. Cham: Springer.

Colin, Samuel and Struyve, Ward (2007): "A Dirac sea pilot-wave model for quantum field theory". *Journal of Physics A* 40, pp. 7309-7341.

Crowther, Karen (2016): *Effective spacetime. Understanding emergence in effective field theory and quantum gravity*. Cham: Springer 2016.

de Broglie, Louis (1928): "La nouvelle dynamique des quanta". In: *Electrons et photons. Rapports et discussions du cinquième Conseil de physique tenu à Bruxelles du 24 au 29 octobre 1927 sous les auspices de l'Institut international de physique Solvay*. Paris: Gauthier-Villars. Pp. 105-132. English translation in G. Bacciagaluppi and A. Valentini (2009): *Quantum theory at the crossroads. Reconsidering the 1927 Solvay conference*. Cambridge: Cambridge University Press. Pp. 341-371.

Deckert, Dirk-André, Esfeld, Michael and Oldofredi, Andrea (2018): "A persistent particle ontology for QFT in terms of the Dirac sea". Forthcoming in *British Journal for the Philosophy of Science*, online first DOI 10.1093/bjps/axx018

Dickson, Michael (2000): "Are there material objects in Bohm's theory?" *Philosophy of Science* 67, pp. 704-710.

Dürr, Detlef, Goldstein, Sheldon, Tumulka, Roderich and Zanghì, Nino (2005): "Bell-type quantum field theories". *Journal of Physics A* 38, pp. R1-R43.

Dürr, Detlef, Goldstein, Sheldon and Zanghì, Nino (2013): *Quantum physics without quantum philosophy*. Berlin: Springer.

Esfeld, Michael (2014): "The primitive ontology of quantum physics: guidelines for an assessment of the proposals". *Studies in History and Philosophy of Modern Physics* 47, pp. 99-106.

Esfeld, Michael, Lazarovici, Dustin, Lam, Vincent and Hubert, Mario (2017): "The physics and metaphysics of primitive stuff". *British Journal for the Philosophy of Science* 68, pp. 133-161.

Everett, Hugh (1957): "'Relative state' formulation of quantum mechanics". *Reviews of Modern Physics* 29, pp. 454-462.

Goldstein, Sheldon, Taylor, James, Tumulka, Roderich and Zanghì, Nino (2005a): "Are all particles identical?" *Journal of Physics A* 38, pp. 1567-1576.

Goldstein, Sheldon, Taylor, James, Tumulka, Roderich and Zanghì, Nino (2005b): "Are all particles real?" *Studies in History and Philosophy of Modern Physics* 37, pp. 103-112.

Goldstein, Sheldon and Teufel, Stefan (2001):"Quantum spacetime without observers: ontological clarity and the conceptual foundations of quantum gravity". In: C. Callender and N. Huggett (eds.): *Physics meets philosophy at the Planck scale*. Cambridge: Cambridge University Press. Pp. 275-289.

Gomes, Henrique, Gryb, Sean and Koslowski, Tim (2011): "Einstein gravity as a 3d conformally invariant theory". *Classical and Quantum Gravity* 28, p. 045005.

Gomes, Henrique and Koslowski, Tim (2013): "Frequently asked questions about shape dynamics". *Foundations of Physics* 43, pp. 1428-1458.

Graham, Daniel W. (2010): *The texts of early Greek philosophy. The complete fragments and selected testimonies of the major Presocratics. Edited and translated by Daniel W. Graham*. Cambridge: Cambridge University Press.

Gryb, Sean and Thébault, Karim P. Y. (2016): "Time remains". *British Journal for the Philosophy of Science* 67, pp. 663-705.

Jackson, Frank (1994): "Armchair metaphysics". In: J. O'Leary-Hawthorne and M. Michael (eds.): *Philosophy in mind*. Dordrecht: Kluwer. Pp. 23-42.

Kiefer, Claus (2007): *Quantum gravity*. Oxford: Oxford University Press. Second edition.

Knox, Eleanor (2018): "Physical relativity from a functionalist perspective". Forthcoming in *Studies in History and Philosophy of Modern Physics*, online first DOI 10.1016/j.shpsb.2017.09.008

Lam, Vincent and Wüthrich, Christian (2018): "Spacetime is as spacetime does". *Studies in History and Philosophy of Modern Physics* 64, pp. 39-51.

Lazarovici, Dustin, Oldofredi, Andrea and Esfeld, Michael (2018): "Observables and unobservables in quantum mechanics: How the no-hidden-variables theorems support the Bohmian particle ontology". *Entropy* 20, pp. 381-397.

Lewis, David (1966): "An argument for the identity theory". *Journal of Philosophy* 63, pp. 17-25.

Lewis, David (1970): "How to define theoretical terms". *Journal of Philosophy* 67, pp. 427-446.

Lewis, David (1972): "Psychophysical and theoretical identifications". *Australasian Journal of Philosophy* 50, pp. 249-258.

Lewis, David (1986): *On the plurality of worlds*. Oxford: Blackwell.

Mach, Ernst (1919): *The science of mechanics: a critical and historical account of its development. Fourth edition. Translation by Thomas J. McCormack*. Chicago: Open Court.

Maudlin, Tim (1995): "Three measurement problems". *Topoi* 14, pp. 7-15.

Millikan, Ruth Garrett (1993): *White Queen psychology and other essays for Alice*. Cambridge (Massachusetts): MIT Press.

Okon, Elias and Sudarsky, Daniel (2014): "Benefits of objective collapse models for cosmology and quantum gravity". *Foundations of Physics* 44, pp. 114-143.

Oriti, Daniele (2014): "Disappearance and emergence of space and time in quantum gravity". *Studies in History and Philosophy of Modern Physics* 46, pp. 186-199.

Pinto-Neto, Nelson and Struyve, Ward (2018): "Bohmian quantum gravity and cosmology". arXiv:1801.03353v1

Rovelli, Carlo (2004): *Quantum gravity*. Cambridge: Cambridge University Press.

Sellars, Wilfrid (1956): "Empiricism and the philosophy of mind". In: H. Feigl and M. Scriven (eds.): *The foundations of science and the concepts of psychology and psychoanalysis*. Minneapolis: University of Minnesota Press. Pp. 253-329.

Tilloy, Antoine (2018): "Binding quantum matter and space-time, without romanticism". *Foundations of Physics* 48, pp. 1753-1769.

Vassallo, Antonio and Esfeld, Michael (2014): "A proposal for a Bohmian ontology of quantum gravity". *Foundations of Physics* 44, pp. 1-18.

Wüthrich, Christian (2017): "Raiders of the lost spacetime". In: D. Lehmkuhl, G. Schiemann and E. Scholz (eds.): *Towards a theory of spacetime theories*. Basel: Birkhäuser. Pp. 297-335.